\documentclass[%
 reprint,
 amsmath,amssymb,
 aps,
]{revtex4}

\usepackage{graphicx}
\usepackage{bm}
\numberwithin{equation}{section}

\newtheorem{theorem}{Theorem}
\newtheorem{claim}{Claim}
\usepackage{hyperref}

\begin{document}

\preprint{JMP}

\title{Weierstrass traveling wave solutions for dissipative BBM equation}

\author{Stefan C. Mancas}
	\email{stefan.mancas@erau.edu}
\author{Greg Spradlin}%
 \email{spradlig@erau.edu}
\author{Harihar Khanal}%
 \email{khana66a@erau.edu}
\affiliation{%
 Department of Mathematics, Embry-Riddle Aeronautical University,\\ Daytona-Beach, FL. 32114-3900, USA
}%

\date{\today}

\begin{abstract}
In this paper the effect of a small dissipation on waves  is included to find exact solutions to the modified BBM equation. Using Lyapunov functions and dynamical systems theory, we prove that when viscosity is added to the BBM equation, in certain regions there still  exist bounded traveling wave solutions in the form of solitary waves, periodic, and  elliptic  functions. By using the canonical form of Abel equation, the polynomial  Appell invariant make the equation integrable in terms of Weierstrass $\wp$  functions. We will use a general formalism based on Ince's transformation to write the general solution  of dissipative BBM in terms of $\wp$  functions, from which all the other known solutions  can be obtained via simplifying assumptions.  Using ODE analysis we show that the traveling wave speed is a  bifurcation parameter that makes transition between different classes of waves.
\begin{description}

\item[PACS numbers]
02.30.Gp, 02.30.Hq, 02.30.Ik, 02.30.Jr
\end{description}
\end{abstract}

\maketitle

\section{Introduction}

In 1895, Korteweg and de Vries, under the assumption of small wave amplitude and large wavelength of inviscid and incompressible fluids, derived an equation for the water waves, now known as the KdV equation \cite{Kor}, which also serves as a justifiable model for long waves in a wide class of nonlinear dispersive systems. KdV has been also used to account adequately for observable phenomena such as the interaction of solitary waves and dissipationless undular shocks. 
For the water wave problem \eqref{KdV} is nondimensionalized since the physical parameters  $|u|=3 \eta /2H$, $x=\sqrt6 x^*/H$, and $t=\sqrt{6g/H}t^*$, where $\eta$ is the vertical displacement of the free surface, $H$ is the depth of the undisturbed water, $x^*$ is dimensional distance and $t^*$ the dimensional time are all scaled into the definition of nondimensional space $x$, time $t$, and water velocity $u(x,t)$. When the physical parameters and scaling factors are appropriately absorbed into the definitions of $u$, $x$ and $t$, the KdV equation is obtained in the tidy form
\begin{equation}\label{KdV}
u_t+u_x+uu_x+u_{xxx}=0.
\end{equation}

Although \eqref{KdV} shows remarkable properties it manifests non-physical properties; the most noticeable being unbounded dispersion relation. It is helpful to recognize that the main difficulties presented by \eqref{KdV} arise from the dispersion term and arise in  the linearized form \cite{Bona}
\begin{equation}\label{lin}
u_t+u_x+u_{xxx}=0.
\end{equation}
First, as BBM noted, when the solution of \eqref{lin} is expressible as a summation of Fourier components in the form $F(k)e^{i (k x-\omega t)}$, the dispersion relation is $\omega(k)=k-k^3$, where $\omega$ is the frequency, and $k$ is the wave number. The phase velocity $C_p=\frac{\omega(k)}{k}=1-k^2$ becomes negative for $k^2>1$, contrary to the original assumption of the forward traveling waves. More significantly, the group velocity $C_g=\frac{d\omega}{d k}=1-3k^2$ has no lower bound. To circumvent this feature, it has been shown by  Peregrine \cite{P}, and later by Benjamin {\it et. al.} \cite{Bona} that  \eqref{KdV} has an alternative format,  which is called the regularized long-wave or  BBM equation
\begin{equation}\label{BBM}
u_t+u_x+uu_x-u_{xxt}=0
\end{equation}
where the dispersion term $u_{xxx}$ is replaced by $-u_{xxt}$ to reflect a bounded dispersion relation. It is contended that \eqref{BBM} is in important respects the preferable model over \eqref{KdV} which is unsuitably posed model for long waves. 

The initial value problem for this equation has been investigated previously by \cite {P,P2,P3}. In \cite{Bona}, the authors   showed that its solutions have better smoothness properties than those of \eqref{KdV}. BBM has the same formal justification and possesses similar properties as that of KdV. 

The linearized version of  \eqref{BBM}  has the  dispersion relation $\omega(k)=\frac{k}{1+k^2}$, according to which both phase and group velocities  $C_p=\frac{1}{1+k^2}$, and $C_g=\frac{1-k^2}{(1+k^2)^2}$ are bounded for all $k$. Moreover, both velocities tend to zero for large $k$, which implies that fine scale features of the solution tend not to propagate.
The preference of \eqref{BBM} over \eqref{KdV} became clear in \cite{Bona}, when the authors attempted to formulate an existence theory for \eqref{KdV}, respective to nonperiodic initial condition $u(x,0)$ defined on $(-\infty,\infty)$. 
 
A generalization to \eqref{BBM} to include a viscous term is provided by the equation
\begin{equation}\label{e0}
u_t+u_x+uu_x-u_{xxt}=\nu u_{xx}
\end{equation}
where $\nu>0$ is transformed  kinematic viscosity coefficient of a liquid is a popular alternative to the Korteweg-de Vries equation, for modeling unidirectional propagation of long waves in systems that manifest nonlinear and dispersive effects.  

The linearized version of  \eqref{e0} has a complex dispersion relation $w(k)=\frac{k-i \nu k^2}{1+k^2}$, so $u(x,t)=e^{ik(x-\frac{t}{1+k^2})}e^{-\frac{\nu k^2}{1+k^2}}$. The real part of the frequency coincides with the bounded dispersion relation of the nondisipative case, while the imaginary part contains the viscosity coefficient $\nu$ which must be positive to have decaying waves. In this case the harmonic solutions exhibit both dispersive and dissipative effects.

  In \cite{Bona}, the equation is recast as an integral equation, and solutions, global in time, are found for initial data in 
$W^{1,2}({\mathbb R})$.  In \cite{BT}, global solutions are found for initial data in $H^s({\mathbb R})$, $s \geq 0$. 
In this article, we study the effect of dissipation on \eqref{BBM}, and henceforth we shall refer to \eqref{e0} as the modified BBM by viscosity.

\medskip

\section{Existence of dissipative solutions}

The existence and stability of solitary waves has been investigated previously by Bona {\it et. al.}, \cite{Bona, Ben, Bona2}. In this section we propose an easy theorem based on dynamical system theory \cite{Baes, Chou}, in which we prove the existence of solutions to \eqref{e0} for both $c<0$, and $c>0$ traveling wave velocities. Also, we will show that the solution tends to the fixed points of the dynamical system $c-1 \pm k$, for any $k$ and large $\zeta$. So, at the both ends of the boundaries the solution does not necessarily decay to zero, but there is flow of energy through the boundaries to the solution and back to the boundaries. 

Theorem 1 below is proven by showing that a certain dynamical system in the plane has a heteroclinic orbit connecting two critical points.    The Lyapunov function $\mathcal{L}$ is used to show that a certain trajectory of the dynamical system is bounded.  The Poincare-Bendixson Theorem \cite{JS} can be applied to show that the trajectory approaches a closed loop or a fixed point.  The arguments in the proof of Theorem 1 show that in the case of a closed loop, $\mathcal{L}$ would have to constant along the loop, which is  impossible by \eqref{e85}.  Therefore the trajectory approaches a fixed point, and in this proof, there is only one fixed point that is possible.

An alternative to the dynamical systems approach to finding
heteroclinic solutions to ordinary differential equations like \eqref{e4},
perhaps with more complicated nonlinearities, are variational methods
and critical point theory. There is an
extensive literature on applying variational methods to these types of equation without the dissipative term. 
See, for example, \cite{BM} and the
references therein.  The inclusion of the $\nu u'$ dissipative term makes this approach more difficult. 
It is necessary to use more complicated function spaces and more
complicated arguments, although this approach was used successfully in \cite{AC}. 

\begin{theorem} \label{thm2}
Let $c \neq 0$ and $k> 0$.  \eqref{e0} has a traveling wave solution of the form $u(x,t) = u(x-ct)=u(\zeta)$ with 
$u(\zeta) \to (c-1)+k$ as $\zeta \to -\infty$ and $u(\zeta) \to (c-1)-k$ as $\zeta \to \infty$.  If $c<0$, then $u(\zeta) < (c-1)+k$ for all real $\zeta$. If $c > 0$, then $u(\zeta) > (c-1)-k$ for all real $\zeta$.
\end{theorem}

{\bf Proof}\\ 
The class of soliton solutions is found by employing the form  of the traveling wave solutions of \eqref{e0} which takes the form of the {\emph {ansatz} 
\begin{equation}\label{4}
u(x,t)=u (\zeta)
\end{equation}
where $\zeta\equiv x-ct$
is the traveling wave variable, and $c$ is a nonzero
translational wave velocity. The substitution of \eqref{4} in \eqref{e0} leads to

\begin{equation}\label{e3}
(1-c)u' + uu' +cu'''=\nu u'',
\end{equation}
where $'=\frac{d}{d\zeta}$. Integrating,
\begin{equation}\label{e4}
(1-c)u + \frac{1}{2}u^2 +cu'' = \nu u' + A
\end{equation}
for some constant $A$.  With $\phi = u$ and $\psi = u'$, \eqref{e4} is equivalent to the 
dynamical system
\begin{equation}\label{e5}
\begin{aligned}
	\phi' &= \psi \\
	 \psi'  &= \frac{\nu}{c}\psi+\frac{c-1}{c}\phi  -\frac{1}{2c}\phi^2 +  \frac{A}{c},\\
\end{aligned}
\end{equation}
or
\begin{equation}\label{e53}
	\mathbf{u}' = \mathbf{F}(\mathbf{u})
\end{equation}
in vector notation.  The fixed points of \eqref{e5} are
\begin{equation}\label{e55}
\mathbf{z}^\pm = \left[ \begin{array}{cc}
                               {\phi^\pm} \\ 
                                 0 
                        \end{array}\right] = 
                 \left[ \begin{array}{cc}
                            {c-1 \pm \sqrt{(c-1)^2+2A}} \\ 
                            0 
                        \end{array}\right]= \left[ \begin{array}{cc} {c-1 \pm k} \\ 0 \end{array}\right],
\end{equation}
for $A=\frac{k^2-(c-1)^2}{2}$. 
We will show that \eqref{e5} has a trajectory from $\mathbf{z}^+$ to $\mathbf{z}^-$.

	First, we will consider the $c < 0$ case.  The linearization of \eqref{e5} at $\mathbf{z}^+$ is
\begin{equation}\label{e65}	
\mathbf{u}' = 
  \left[ \begin{array}{c}
            \phi' \\
            \psi'
         \end{array}\right] =    
\left[ \begin{array}{cc}
                    0 & 1  \\ 
               -\frac{\sqrt{(c-1)^2+2A}}{c} & \frac{\nu}{c} 
                      \end{array}\right] 
              \left[ \begin{array}{c}
                   \phi - \phi^+ \\ 
                    \psi
                      \end{array}\right]   	\equiv
                M\left[ \begin{array}{c}
                           \phi - \phi^+ \\ 
                           \psi
                           \end{array}\right].
\end{equation}
The determinant of $M$ is negative, so $\mathbf{z}^+$ is a saddle point, and $M$ has eigenvalues $\lambda^\pm$ with 
$\lambda^- < 0 < \lambda^+$. 

 The unit eigenvector corresponding to $\lambda^+$ and pointing to the left is
\begin{equation}\label{e66}
\mathbf{v} = \frac{\left[ \begin{array}{c}
                               -1 \\
                               -\lambda^+
                           \end{array}\right]}{\sqrt{1+(\lambda^+)^2}}.  
\end{equation}

Let $\mathbf{g(\zeta)}=\left[ \begin{array}{c}
                               g_1(\zeta) \\
                               g_2(\zeta)
                           \end{array}\right]$ be a solution of \eqref{e5}
with $\mathbf{g}(\zeta) \to \mathbf{z}^+$ and 
\begin{equation}\label{e67}
\Big| \frac{\mathbf{g}(\zeta) -\mathbf{z}^+}{|\mathbf{g}(\zeta) -\mathbf{z}^+ |} - \mathbf{v} \Big| \to 0
\end{equation}
as $\zeta \to -\infty$.
We will show that $\mathbf{g}(\zeta) \to \mathbf{z}^-$ as $\zeta \to \infty$ and $g_1(\zeta) < \phi^+$ for all real $\zeta$.
Define
\begin{equation}\label{e7}
P_3(\phi)= \frac{1}{6}\phi^3 + \frac{1}{2}(1-c)\phi^2-A\phi.
\end{equation}
$P_3$ is increasing on $(-\infty, \phi^-]$, decreasing on $[\phi^-, \phi^+]$, and increasing on $[\phi^+, \infty)$.  Define
\begin{equation}\label{e8}
\begin{aligned}
{\mathcal L}(\mathbf{u}) = & P_3(\phi)-P_3(\phi^+) + \frac{1}{2}c \psi^2 \\
= &							\frac{1}{6}\phi^3+\frac{1}{2}(1-c)\phi^2-A\phi -P_3(\phi^+) + \frac{1}{2}c \psi^2.
							\end{aligned}
\end{equation}
Then for all $\zeta$,
\begin{equation}\label{e85}
\begin{aligned}
	\frac{d}{d\zeta}\mathcal{L}(\mathbf{g}(\zeta)) &= \Big(\frac{1}{2}g_1^2 + (1-c)g_1 - A\Big)g'_1 +cg_2 g'_2 =\\
	        &= \Big(\frac{1}{2}g_1^2 + (1-c)g_1-A\Big)g_2 + \\
	        &\qquad cg_2\Big(\frac{c-1}{c}g_1 -\frac{1}{2c}g_1^2 + \frac{\nu}{c}g_2 + \frac{A}{c}\Big) = \nu g_2^2 \geq 0.\\
\end{aligned}
\end{equation}
Since $\mathcal{L}(\mathbf{z}^+) = 0$ then $\mathcal{L}(\mathbf{g}(\zeta)) > 0$ for all real $\zeta$.

	Let $\phi^{--} < \phi^-$ with
\begin{equation}\label{e9}
											P_3(\phi^{--}) < P_3(\phi^+),
\end{equation}
and let
\begin{equation}\label{e10}
						B = P_3(\phi^-) - P_3(\phi^+) = \max\{P_3(\phi) \mid \phi^{--} \leq \phi \leq \phi^+ \} - P_3(\phi^+).
\end{equation}
Let $L >0$ be large enough so that
\begin{equation}\label{e11}
			\frac{1}{2}c L^2 + B < 0,
\end{equation}
and define
\begin{equation}\label{e12}
					R = (\phi^{--} , \phi^+) \times (-L,L).
\end{equation}	  
We claim that $\mathbf{g}(\zeta) \in R$ for all real $\zeta$.  As $\zeta \to -\infty$, $\mathbf{g}(\zeta) \to \mathbf{z}^+$ along the direction 
$[1, \lambda^+]^T$.  Therefore $\mathcal{L}(\mathbf{g}(\zeta)) > \mathcal{L}(\mathbf{z}^+) = 0$ for all real $\zeta$, and there exists 
$\zeta` \in \mathbb{R}$ such that $\mathbf{g}(\zeta) \in R$ for all $\zeta < \zeta`$.  

So, it suffices to show that $\mathcal{L} \leq 0$ on $\partial R$.  Let $\mathbf{u} \in \partial R$.
If $\phi = \phi^+$, then $\mathcal{L}(\mathbf{u}) = \frac{1}{2}c \psi^2 \leq 0$.  Similarly, if $\phi = \phi^{--}$, then $\mathcal{L}(\mathbf{u}) = P_3(\phi^{--}) - P_3(\phi^+) +\frac{1}{2}c \psi^2 < 0$.  If $\phi^{--} \leq \phi \leq \phi^+$ and $\psi = L$, 
then $\mathcal{L}(\mathbf{u}) = P_3(\phi)-P_3(\phi^+) + \frac{1}{2}c L^2 \leq B + \frac{1}{2}c L^2 < 0$.  Similarly, if $\phi^{--} \leq \phi \leq \phi^+$ and $\psi = -L$, $\mathcal{L}(\mathbf{u}) < 0$.

Define 
\begin{equation}\label{e13}
	L_\infty = \lim_{\zeta \to \infty} \mathcal{L}(\mathbf{g}(\zeta)).
\end{equation}	
$L_\infty$ exists and is finite because $\mathcal{L}(\mathbf{g}(\zeta))$ is a non-decreasing function of $\zeta$, $\mathcal{L}$ is bounded on $R$, and $\mathbf{g}(\zeta) \in R$ for all real $\zeta$.  Let $C$ be the $\omega$-limit set of $\mathbf{g}$, i.e., 
\begin{equation}\label{e14}
C = \{\mathbf{u} \mid \exists (\zeta_m) \subset \mathbb{R} \mbox{ with } \lim_{m \to \infty} \zeta_m = \infty, \ \mathbf{g}(\zeta_m) \to \mathbf{u}\}
\end{equation}
$C$ is nonempty because $\mathbf{g}(\zeta) \in R$ for all real $\zeta$ and $R$ is bounded.  $C$ is compact and connected, and $\mathcal{L} = L_\infty$ on $C$.  We must prove $C = \{\mathbf{z}^-\}$.

	Suppose $C$ contains a point $\mathbf{w}$ that is not on the $\phi$-axis.  Then $\mathbf{w}$ is not a stationary point of \eqref{e5}, and $C$ contains another point $\mathbf{y}$.  Let $\delta > 0$ with $\delta < \min(|\mathbf{w} - \mathbf{y}|, |w_2|)/2$.  Then $\mathbf{g}$ crosses the annulus 
	$B_\delta(\mathbf{w}) \setminus B_{\delta/2}(\mathbf{w})$ infinitely many times, and there exist sequences $(\zeta_m), (\zeta`_m)$ with $\zeta_m \to \infty$ as $m \to \infty$, $\zeta_m < \zeta`_m < \zeta_{m+1}$ for all $m$, $|\mathbf{g}(\zeta_m) - \mathbf{w}| = \delta/2$, $|\mathbf{g}(\zeta`_m) - \mathbf{w}| = \delta$ and $\delta/2 < |\mathbf{g}(\zeta) - \mathbf{w}| < \delta$ for all $m$ and all $\zeta \in (\zeta_m, \zeta`_m)$. 
	
	Let $K = \sup(|\mathbf{F}(\mathbf{u})| \mid \mathbf{u} \in B_\delta(\mathbf{w})) < \infty$, where $\mathbf{F}$ is from \eqref{e53}.  Then $\zeta`_m - \zeta_m \geq \delta/(2K)$ for all $m$.  
For $\zeta \in (\zeta_m, \zeta`_m)$, 
\begin{equation}\label{e15}
\frac{d}{d\zeta} \mathcal{L}(\mathbf{g}(\zeta)) \geq \nu g_2^2(\zeta) \geq \frac{\nu w_2^2}{4}> 0,
\end{equation}	
so 
\begin{equation}\label{e16}
\begin{aligned}
\mathcal{L}(\mathbf{g}(\zeta`_m))&- \mathcal{L}(\mathbf{g}(\zeta_m)) =
	\int_{\zeta_m}^{\zeta`_m}\frac{d}{d\zeta} \mathcal{L}(\mathbf{g}(\zeta))\, d\zeta \geq  \\	
	&\int_{\zeta_m}^{\zeta`_m} \frac{\nu w_2^2}{4}\, d\zeta =
		\frac{\nu(\zeta`_m - \zeta_m)w_2^2}{4} \geq \frac{\delta \nu w_2^2}{8K}
\end{aligned}
\end{equation}	  
for all $m$.  Therefore,
\begin{equation}\label{e17}
\begin{aligned}
L_\infty &= \mathcal{L}(\mathbf{g}(\zeta_1)) + (L_\infty - \mathcal{L}(\mathbf{g}(\zeta_1)) \geq \\
         &\geq \mathcal{L}(\mathbf{g}(\zeta_1))+ 
         \sum_{m=1}^\infty \mathcal{L}(\mathbf{g}(\zeta`_m))- \mathcal{L}(\mathbf{g}(\zeta_m)) \geq \\
         &\geq \mathcal{L}(\mathbf{g}(\zeta_1)) + \sum_{m=1}^\infty \frac{\delta \nu w_2^2}{8K} = \infty.
\end{aligned}
\end{equation}
But $L_\infty$ is finite.  This is a contradiction.  

Therefore, $C$ is a subset of the $\phi$-axis.  Since $C$ is connected and compact, 
$C = [\phi_1, \phi_2] \times \{0\}$ for some $\phi^{--} < \phi_1 \leq \phi_2 < \phi^+$.  Recall that $\mathcal{L} = L_\infty$ on $C$.  $\mathcal{L}$ is not constant on any segment of the $\phi$-axis of positive length.  So $C$ is a single point, which is a stationary point of \eqref{e5} in $R$.  There is only one such point, $\mathbf{z}^-$.

Now consider the case $c > 0$.  Let $k > 0$.  Since $-c < 0$, we have seen that there exists $g$ solving \eqref{e3} with $c$ replaced by $-c$, that is, 

\begin{equation}\label{e18}
(1+c)g' + gg' -cg''' - \nu g'' = 0,	
\end{equation}	
with $g(\zeta) \to (-c-1)+k$ as $\zeta \to -\infty$, $g(\zeta) \to (-c-1)-k$ as $\zeta \to \infty$, and $g(\zeta) < (-c-1) + k$ for all real $\zeta$.  

Define $u: \mathbb{R} \to \mathbb{R}$ by $u(\zeta) = -g(-\zeta)-2$. Then $g(\zeta) = -u(-\zeta)-2$, $g'(\zeta) = u'(-\zeta)$, $g''(\zeta) = -u''(-\zeta)$, and $g'''(\zeta) = u'''(-\zeta)$.  Substituting into \eqref{e18} yields
\begin{equation}\label{e19}
\begin{aligned}
 0 &= (1+c)u'(-\zeta) + (-u(-\zeta)-2)u'(-\zeta) -cu'''(-\zeta) + \nu u''(-\zeta) \\
   &= (c-1)u'(-\zeta) - u(-\zeta)u'(-\zeta) - cu'''(-\zeta) + \nu u''(-\zeta) \\
   &= -[(1-c)u'(-\zeta) + u(-\zeta)u'(-\zeta) + cu'''(-\zeta) -\nu u''(-\zeta)].
\end{aligned}
\end{equation}  
Therefore $u$ satisfies \eqref{e3}.  By definition of $u$, 
\begin{equation}\label{e20}
\begin{aligned}
\lim_{\zeta \to -\infty} u(\zeta) &= \lim_{\zeta \to -\infty} ( -g(-\zeta)-2) = \lim_{\zeta \to \infty} (-g(\zeta)-2) \\
				&= -2 -\lim_{\zeta \to \infty} g(\zeta) = -2 - ((-c -1) -k)\\
				& = (c-1)+k,
\end{aligned}
\end{equation}
and
\begin{equation}\label{e21}
\begin{aligned}
\lim_{\zeta \to \infty} u(\zeta) &= \lim_{\zeta \to \infty} ( -g(-\zeta)-2) = \lim_{\zeta \to -\infty} (-g(\zeta)-2)  \\
				&= -2 -\lim_{\zeta \to -\infty} g(\zeta) = -2 - ((-c -1) + k) \\
				&= (c-1)-k.
\end{aligned}
\end{equation}
For all real $\zeta$, $g(\zeta) < (-c-1)+k$, so for all real $\zeta$,
\begin{equation}\label{e22}
u(\zeta) = -g(-\zeta) - 2 > -[(-c-1)+k] - 2 = (c-1) - k.
\end{equation}
Theorem~\ref{thm2} is proven.	

\medskip
\section{Exact solutions via $\wp$ functions}
\subsection{Abel's equation}
The importance of Abel's equation in its canonical forms stems from the fact that its integrability leads to closed form solutions to a general nonlinear ODE of the form
\begin{equation}\label{n1p}
u_{\zeta \zeta}+q_0(u) u_{\zeta}  +q_2(u)=0~.
\end{equation}
This can be expressed by the following lemma \cite{Man}.

\begin{quote}
{\bf Lemma 1}: Solutions to a general second order ODE of type (\ref{n1p}) may be obtained via the solutions to Abel's equation (\ref{n2p}), and vice versa using the following relationship
%
\begin{equation}\label{n10}
u_{\zeta}=\eta(u(\zeta))~.
\end{equation}
\end{quote}

{\bf Proof}: To show the equivalence, one just need the simple chain rule
\begin{equation}\label{eq1}
u_{\zeta\zeta}=\frac{d \eta}{d u}u_{\zeta}=\eta \frac{d \eta}{d u}
\end{equation}
which turns (\ref{n1p}) into the second kind of Abel equation
\begin{equation}\label{eq2}
\frac{d \eta}{du}\eta+q_0(u)\eta+q_2(u)=0.
\end{equation}
Moreover, via transformation of the dependent variable
\begin{equation}\label{eq3}
\eta(u)=\frac{1}{y(u)}
\end{equation}
(\ref{eq2}) becomes the Abel first kind without free or linear terms
\begin{equation}\label{n2p}
\frac{dy}{du}=q_0(u)y^2+q_2(u)y^3.
\end{equation}
Comparing \eqref{n1p} with \eqref{e4} we identify the constant, and quadratic polynomial coefficients
				\begin{equation}	\label{Ab}
					\begin{aligned}
					q_0(u)& =-\frac{\nu}{c}\\
					q_2(u)&=\frac{u^2}{2c}+\frac{1-c}{c}u-\frac{A}{c}
				\end{aligned}
\end{equation}
The need of dissipation is imperative since without the dissipation term $q_0=0$, the Abel'e equation \eqref{n2p} becomes separable, as we will see next.

Progress of integration of \eqref{n2p} is based on the linear transformation $v=\int{q_0(u) du}=-\frac\nu c u$, which allows us to write Abel's equation \eqref{n2p} in canonical form 
	
	\begin{equation}\label{Ab2}
				\frac {dy}{d v }=y^2+g(v)y^3.
				\end{equation}
				$g(v)$ is the Appell's invariant and is the quadratic polynomial
				\begin{equation}\label{Ap}
				g(v)=-\frac{c^2}{2 \nu^3}v^2+\frac{c(1-c)}{\nu^2}v+\frac{A}{\nu} = a_2v^2+a_1v+a_0
				\end{equation}
				
By Lemke's transformation \cite{Lem}
				\begin{equation}
				y=-\frac{1}{z}\frac{d z}{d v}
				\end{equation}
\eqref{Ab2} can be written as a second-order differential equation with a preserved invariant $g(v)$.
\begin{equation}\label{sys}
					z^2\frac{d^2 v}{d z^2}+g(v)=0
					\end{equation}	
					
					\subsection{No viscosity}
If $\nu=0$ \eqref{e4} becomes

\begin{equation}\label{61}
u_{\zeta \zeta}=-\frac{u^2}{2c}-\frac{1-c}{c}u+\frac{A}{c} 
\end{equation}
The classical solutions obtained before \cite {Ben} by assuming $A=0$ are

\begin{align}\label{62}
u(x,t)&=3(c-1) \mathrm{Sech}^2\Big[{\frac{\sqrt{(c-1)/c}}{2}(x-c t)}\Big] \,\, \mathrm{if} \,\, c>1  \\ 
u(x,t)&=-\frac{3c}{1+c} \mathrm{Sec}^2\Big[{\frac{\sqrt {c}}{2}\Big (x-t/(1+c)\Big)}\Big] \,\, \mathrm{if} \,\, 0<c<1 \notag
\end{align}
see left and middle panels of FIG. 1. The solitary waves that move with velocity $c>1$ will be called fast waves, while the periodic solutions that travel with velocity $0<c<1$ slow waves.  

In \eqref{n2p} if let $\nu=0$, then the Abel's equation
\begin{equation}
\frac{dy}{du}=q_2(u)y^3
\end{equation}
is separable and leads to the \emph{elliptic algebraic} equation 
\begin{equation}\label{1001}
\eta ^2=-\frac{u^3}{3c}-\frac{1-c}{c}u^2+\frac{2A}{c}-2D.
\end{equation} 

Using again the substitution $u_{\zeta}=\eta$ \eqref{1001} becomes the \emph{elliptic differential} equation 
\begin{equation}\label{aaa}
u_{\zeta} ^2=-\frac{u^3}{3c}-\frac{1-c}{c}u^2+\frac{2A}{c}u-2D.
\end{equation}
For verification, if one takes $\frac{d}{d\zeta}$ of the above, we recover \eqref{61}. 

Using a linear transformation \cite{Byrd} p. 311 of the dependent variable $u=-\sqrt[3]{12c}\hat{u}-(1-c)$, \eqref{aaa} can be written in standard form
\begin{equation}\label{sol3}
\hat u_{\zeta}^2=4 \hat{u}^3-g_2 \hat{u}-g_3,
\end{equation}  with solution
\begin{equation}
\hat u(\zeta)=\wp(\zeta, g_2,g_3)
\end{equation} and invariants
\begin{align}
		g_2=&\sqrt[3]{\frac{12}{c^2}}(c-1)^2>0\\
		g_3=&\frac{2(1-c)^3}{3c}.
		\end{align}
The limiting cases are obtained when $A=0, D=0$ to yield back to \eqref{62}.  We will not discuss here this reduction to hyperbolic or trigonometric functions, since this was done extensible before by many authors, but instead we refer the reader to \cite{Nickel, Ablo} for the reduction and classification that depends on the signs of the invariants. 

Instead, we will analyze the novel case of the solutions that travel with a critical velocity $c=1$. At the boundary between the fast and slow waves we will encounter periodic solutions in terms of rational combination of Jacobian elliptic functions.

If $c=1$ and $A=0$, eq. \eqref{61} becomes
\begin{equation}\label{el1}
\frac{u^2}{2}+u''=0.
\end{equation}
By multiplying by $u_{\zeta}$ and integrating once we obtain 
\begin{equation}\label{el2}
u_{\zeta}^2+\frac{u^3}{3}=\frac{B^3}{3},
\end{equation}
where $B$ is some nonzero constant of integration. In \eqref{1001} if we let $D=-\frac{B^3}{6}$, we will also arrive to \eqref{el2}.
Now let's use the substitution $\mu^2=B-u$, where $\mu=\mu(\zeta)$. Hence \eqref{el2} is
\begin{equation}\label{el3}
\mu_{\zeta}= \pm\frac {\sqrt 3}{6} \sqrt{\mu^4-3B\mu^2+3B^2}.
\end{equation} 
Moreover, let's assume $\mu(\zeta)=\sqrt[4]{3} \sqrt{B}z(\zeta)$, then \eqref{el3} becomes
\begin{equation}\label{el4}
z_{\zeta}= \pm\frac {\sqrt B}{2 \sqrt[4]{3}} \sqrt{z^4-\sqrt{3}z^2+1}.
\end{equation} 
This differential equation will be solved using Jacobian elliptic functions.\\ 

\begin{claim}if $Z=1+2z^2 \cos {2\alpha}+z^4$, then $$u(x,k)=\int^{x}_{0}\frac{dz}{\sqrt{Z}}=\frac 12 {\rm sn}^{-1}\frac{2x}{1+x^2},$$ with $k=\sin{\alpha}$, \cite{Bowman} p. 91.
\end{claim}

{\bf Proof}\\ Putting $z=\tan{\theta}$, we find $$u(x,k)=\int^{\tan^{-1}x}_{0}\frac{d \theta}{\sqrt{1-\sin^{2}\alpha \sin^{2}{2 \theta}}},$$ followed by  $y=\sin{2\theta}$, which will give us
$$u(x,k)=\frac 12\int^{\frac{2x}{1+x^2}}_{0}\frac{d y}{\sqrt{(1-y^2)(1-k^2 y^2)}},$$ where the integrand is the elliptic integral of the first kind.}
Therefore, solving \eqref{el4}, we obtain the solution in $z$
\begin{equation}\label{el5}
\frac{2 z}{1+z^2}=\pm {\rm sn}\big(\frac{\sqrt{B}}{\sqrt[4]{3}}\zeta,k\big),
\end{equation}
where $k=\sin{\frac{5 \pi}{12}}=\frac{\sqrt{3}+1}{2 \sqrt{2}}$ is the modulus of the Jacobian elliptic function. It follows that in $\mu$ we will have
\begin{equation}\label{el6}
\frac{2 a \mu }{a^2+\mu^2}=\pm {\rm sn}\big(\frac{\sqrt3}{3} a\zeta,k\big),
\end{equation}
where $a=\sqrt B\sqrt[4]{3}$.
By solving the quadratic, we have 
\begin{equation}\label{el7}
\mu=\pm a\frac{1\pm {\rm cn}\big(\frac{\sqrt 3}{3}a\zeta,k\big)}{{\rm sn}\big(\frac{\sqrt 3}{3}a\zeta,k\big)}.
\end{equation}

Therefore, the solution to the BBM eq. without viscosity and velocity $c=1$ is 
\begin{equation}\label{el8}
\begin{aligned}
u(x,t)=& B\Big(1-\sqrt{3}\Big(\frac{1\pm {\rm cn}\big(\frac{\sqrt B}{\sqrt[4]{3}}(x-t),k\big)}{{\rm sn}\big(\frac{\sqrt B}{\sqrt[4]{3}}(x-t),k\big)}\Big)^2\Big)\\
=& B\Big(1-\sqrt{3
}\frac{1\mp {\rm cn}\big(\frac{\sqrt B}{\sqrt[4]{3}}(x-t),k\big)}{1 \pm {\rm cn}\big(\frac{\sqrt B}{\sqrt[4]{3}}(x-t),k\big)}\Big).
\end{aligned}
\end{equation}
These critical solutions are 
$2K$ periodic with $K$ given by the complete elliptic integral
\begin{equation} 
\frac K2=\int_0^{1}\frac{dz}{\sqrt{z^4-\sqrt{3}z^2+1}}=2.76806,
\end{equation}
see right panel of FIG. 1. The same solution was previously obtained by Cornejo-P\'erez and  Rosu, using a factorization technique \cite{Rosu}.

The expressions \eqref{62} and \eqref{el8}, describe the whole class of solitary wave solutions, with spectrum $c \in(0,\infty)$ which is in concordance with \cite{Ben}. They present periodic and solitary wave solutions to \eqref{BBM}, with no restrictions on the constant $A$, which may be advantageous when adapting  results to physical problems. 

{\bf Remark 1}:  all the solutions obtained using simplifying assumptions $\nu=0$, $A=0$ are particular cases of general solutions of \eqref{e4} in terms of Weierstrass elliptic functions $\wp$, as we will see next.
					
					\subsection{Viscosity present, $\nu>0$}
					
Consider \eqref{Ab2} in the form of non-autonomous eq $F(y,y_v,v)=0$. Painlev\'e proved in 1887 that all integrals of non-autonomous equations do not have movable singular points, but only poles and fixed algebraic singularities. Poincare \cite{Poi} proved in 1885 that any non-autonomous equation of genus $p=0$ is reducible to Riccati eq, while if $p=1$ is integrable via Weierstrass $\wp$ functions, after a linear fractional transformation. Since Riccati equation can be easily obtained from a linear ODE and the Appel invariant $g(v)$ has no singularities, one can conclude that the closed form solutions of \eqref{sys} will not have movable singular points, but poles. Since $v(z)$ has only poles of order two, then the  solution to \eqref{sys}, must be written in terms of Weierstrass $\wp$ functions via the transformation
					\begin{equation}
					v=E z^{p}\omega(z) +F,
					\end{equation}
					see Ince \cite{Ince} p. 431. 
					By substituting this ansatz in \eqref{sys} with $E, F$ constants, then if $p=\frac 25$, $\omega$ will satisfy the elliptic equation 
					
\begin{equation}  \label{eq7}
\omega'^2=4 \omega^3-g_2\omega-g_3
\end{equation} with elliptic invariants $g_2, g_3$. 
					
					Moreover, $z$ is an exponential function, because it satisfies 
					\begin{equation}
					\frac{dv}{dz}=-\frac 1 {yz}
					\end{equation}
					which leads to 
					\begin{equation}
					\frac{\nu}{c}d \zeta=\frac{dz}{z}.
					\end{equation}

The $\wp$ solutions can be combined into the general substitution
\begin{equation}\label{98}
u(\zeta)=\sigma-g(\zeta), 
\end{equation}
where  $g(\zeta)=e^{-n \zeta}\Omega(\zeta)$. $\sigma$, and $n$ are constants which are related to $A$ and $p$. 

By substituting \eqref{98} into \eqref{e4} we obtain the ODE
\begin{equation}\label{100}
g''-\frac{\nu}{c}g'-\frac{1}{2c}g^2+\frac{1-c+\sigma}{c} g=0.
\end{equation}
The free term was eliminated by setting $A=\frac{\sigma^2}{2}+\sigma(1-c)$. Since  $g(\zeta)=e^{-n \zeta} \Omega(\zeta)$, \eqref{100} becomes 
\begin{equation}\label{101}
\Omega''-\big(2n+\frac {\nu}{c}\big)\Omega'+\big(n^2+\frac{n \nu}{c}+\frac{1-c+\sigma}{c}\big)\Omega=\frac{1}{2c}e^{-n\zeta}\Omega^2
\end{equation}
Finally, let $\Omega(\zeta)=\omega(z(\zeta))$, and by chain rule with $'=\frac{d}{d \zeta}$ and $\dot{}=\frac{d}{dz}$, \eqref{101} is
\begin{equation}\label{102}
(z')^2  \ddot \omega+\Big(z''-\big(2n+\frac{\nu}{c}\big)z'\Big)\dot \omega+\Big(n^2+\frac{1-c+\sigma+n \nu}{c}\Big)\omega=\frac{1}{2c}e^{-n \zeta}\omega^2
\end{equation}
By letting 
\begin{equation}\label{103}
z''-\big(2n+\frac{\nu}{c}\big)z'=0
\end{equation} we obtain
\begin{equation}\label{104}
 z'(\zeta)=c_1e^{(2n+\frac{\nu}{c})\zeta}
 \end{equation}
 We also choose $\sigma=-(n^2c+n \nu +1-c)$ which cancels  the linear term in \eqref{102}. 
 
 We are left to solve
 \begin{equation}\label{105}
z'^2 \ddot \omega-\frac{1}{2c}e^{-n \zeta}\omega^2=0 
\end{equation}
 subject to \eqref{104}. Set  $n=-p \frac {\nu}{c}=-\frac{2 \nu}{5c}$, then 
 \begin{equation}\label{140}
\begin{aligned}
	\sigma &=\frac{14 \nu^2}{25 c}+c-1\\
	 A  &= -\frac{\sigma^2}{2}+\frac{14 \nu^2 \sigma}{25c}.\\
\end{aligned}
\end{equation}

By substituting \eqref{104} into \eqref{105}, we obtain

\begin{equation}\label{106}
\ddot \omega=\frac{1}{2cc_1^2}\omega^2.
\end{equation}

Letting $c_1=\frac {1}{2  \sqrt{ 3c}}$, we arrive at the elliptic equation
 \begin{equation}\label{107}
\ddot \omega=6\omega^2
\end{equation}
 which by multiplication by $\dot \omega$ and integration becomes
 \begin{equation}\label{108}
\dot \omega^2=4\omega^3-g_3.
\end{equation}
Its solution is
\begin{equation}\label{108p}
\omega(z)=\wp(z+c_3,0,g_3)
\end{equation}

with invariants $g_2=0$, and $g_3$. If $g_3=1$, this is the equianharmonic case, see \cite{Ablo} p.652.
 
$z(\zeta)$ is found by integrating \eqref{104} to get
\begin{equation}\label{109}
z(\zeta)=c_2+\frac{5 \sqrt{3c}}{6\nu}e^{\frac{\nu \zeta}{5c}}
\end{equation}

\begin{equation}\label{110}
\Omega(\zeta)=\omega(z)=\wp(c_4+ \frac{5 \sqrt{3c}}{6\nu}e^{\frac{\nu \zeta}{5c}},0,g_3),
\end{equation}
where $c_4=c_2+c_3$, and $g_3$ are two integration constants. Using all of the above, together with \eqref{140} we obtain 
\begin{equation}\label{141}
\sigma=\frac{14 \nu^2}{25 c}\pm \sqrt{\Big(\frac{14 \nu^2}{25 c}\Big)^2-2A}
\end{equation}
Then,   the general solution to \eqref{eq7} is
\begin{equation}\label{sol}
u(\zeta)=\frac{14 \nu^2}{25 c}\pm \sqrt{\Big(\frac{14 \nu^2}{25 c}\Big)^2-2A(\nu,c)} -e^{\frac{2 \nu \zeta}{5c}}\wp(c_4+\frac{5 \sqrt{3c}}{6\nu}e^{\frac{\nu \zeta}{5c}},0,g_3),
\end{equation}
see FIGS. 2,3.
Once we fix the velocity $c$, and dissipation $\nu$, then the constants $\sigma$, and $A(\nu,c)$ are obtained using \eqref{140},  which subsituted back into \eqref{sol}, will give us the solution to \eqref{e0}. The two  integration constants that depend on the boundary conditions are $c_4$, and $g_3$. It is crucial to notice that the fixed points of \eqref{e5} depend on the velocity $c$ and dissipation $\nu$ via the constant $A$, see \eqref{e55} 
Finally, it is worth mentioning that the solution in terms of $\wp$ functions is not unique. One can see that there is at least one other similar form previously obtained by Porubov, via the Carnevalle method and B$\ddot{a}$cklund transformation \cite{Porubov, Kud}.\\

{\bf Remark 2:}
Letting $\nu=0\rightarrow n=0$, $c_1=1$ and $c=1\rightarrow \sigma=0$ in \eqref{106}, and since $\omega=g$, but $f=-g$, then $\omega=-f$ so \eqref{108} becomes \eqref{el1}, with solution given by \eqref{el8}, see FIG. 1 right. \\

{\bf Remark 3:}
Letting $\nu=0\rightarrow n=0$, and $\sigma=c-1\ne0$, then according to \eqref{104}, $z'=c_1$. Since $\omega=g$, then \eqref{102} is actually \eqref{61}, and hence we recover the slow and fast waves solutions \eqref{62}, see FIG. 1 left and middle.\\

{\bf Remark 4:} If the second invariant $g_3=0$ in \eqref{108}, then by integrating \eqref{108}
\begin{equation}\label{111}
\omega(z)=\frac{1}{(c_5\pm z)^2}.
\end{equation}
Using \eqref{109}  in \eqref{111} we obtain the kink solution

\begin{equation}\label{112}
u(\zeta)=\frac{14 \nu^2}{25 c}\pm \sqrt{\Big(\frac{14 \nu^2}{25 c}\Big)^2-2A}-\frac{e^{\frac{2 \nu \zeta}{5 c}}}{\Big(c_6\pm \frac{5 \sqrt{3c}}{6\nu}e^{\frac{\nu \zeta}{5c}} \Big)^2},
\end{equation}
see FIG. 4.

An analysis when a $\wp$ function tends to a smooth kink can be found in \cite{Samsonov}.

{\bf Remark 5:} 
If $A=0 \rightarrow c-1=\pm k=\pm\frac{14 \nu^2}{25c}$, and one selects the lower branch of the radical, we obtain
\begin{equation}\label{sol2}
\begin{aligned}
u(\zeta)= & -e^{\frac{2 \nu \zeta}{5c}}\wp(c_4+\frac{5}{\nu}\sqrt{\frac{c}{3a}}e^{\frac{\nu \zeta}{5c}},0,g_3) \,\, \mathrm{if} \,\,g_3 \ne0\\
u(\zeta)= & -\frac{e^{\frac{2 \nu \zeta}{5 c}}}{\Big(c_6\pm \frac{5 \sqrt{3c}}{6\nu}e^{\frac{\nu \zeta}{5c}} \Big)^2} \,\, \mathrm{if} \,\,g_3 =0.\\
\end{aligned} 
\end{equation}

\section{Stability of the viscous waves}

In this section we consider the two-mode dynamical system of \eqref{e5} with equilibrium points in the phase plane $(\phi,\psi)$ at $(0,0)$ and $(c-1\pm \sqrt{(c-1)^2+2A},0)$, where $A=A(\nu,c)$ is obtained from \eqref{140}. All the equilibrium points lie only in the plane $\psi=0$, in the $(\phi,\psi,c)$ space. The bifurcation curves are given by $\phi\big(\phi-(c-1)\mp \sqrt{(c-1)^2-2A}\big)=0$, which are two constant lines.
Near the origin $(\phi^-,0)=(0,0)$
\begin{eqnarray}\label{17}
\left[\begin{array}{cc}
\phi'\\
\psi'\\
\end{array}\right]=\left[\begin{array}{cc}
0& 1\\
-\frac{1-c}{c} & \frac{\nu}{c}\\
\end{array}\right]\left[\begin{array}{cc}
\phi\\
\psi\\
\end{array}\right].
\end{eqnarray}
Following standard methods of phase-plane analysis the characteristic polynomial of the Jacobian matrix of \eqref{17} evaluated at the fixed point $(\phi^-,0)$ is
\begin{equation} \label{charac1}
g_0(\lambda)=\lambda^2-p_0\lambda +q_0=0,
\end{equation}
where $p_0=\frac{\nu}{c}$, and $q_0=\frac{1-c}{c}$.

Since $\nu>0$, $c>0$, then $p_0>0$, hence the origin is \underline{unstable}. Also, putting $\Delta=p_0^2-4q_0=\frac{\nu^2-4c(1-c)}{c^2}$,
we have the following cases:
\begin{itemize}
\item[(i)] $0<c<1$, gives $q_0>0$. If $\nu>2\sqrt{c(1-c)}>0$, the origin is unstable \emph{node}; if $0<\nu<2\sqrt{c(1-c)}$, the origin is unstable \emph{spiral}, see FIG. 5, 
\item[(ii)] $c>1$, gives $q_0<0\Rightarrow \Delta >0$, hence the origin is a \textit{saddle} point, see FIG. 6.
\end{itemize}

Near the secondary fixed point $(\phi^+,0)=(c-1\pm \sqrt{(c-1)^2+2A} ,0)$
\begin{eqnarray}\label{99}
\left[\begin{array}{cc}
\phi'\\
\psi'\\
\end{array}\right]=\left[\begin{array}{cc}
0& 1\\
\frac{1-c}{c} & \frac{\nu}{c}\\
\end{array}\right]\left[\begin{array}{cc}
\phi\\
\psi\\
\end{array}\right].
\end{eqnarray}
The characteristic polynomial of the Jacobian matrix of \eqref{99} evaluated at the fixed point $(\phi^+,0)$ is
\begin{equation} \label{charac2}
g_1(\lambda)=\lambda^2-p_1\lambda +q_1=0,
\end{equation}
where $p_1=p_0=\frac{\nu}{c}$, and $q_1=-q_0=-\frac{1-c}{c}$.

Since $p_1=p_0$, the second fixed point is also \underline{unstable}, and moreover we have the cases:

\begin{itemize}
\item[(i)] $0<c<1$, gives $q_1<0\Rightarrow \Delta>0$, hence the fixed point is a \textit{saddle} point, see FIG. 5,
\item[(ii)] $c>1$, gives $q_1>0$.  if $\nu>2\sqrt{c(c-1)}$,  we have an unstable \emph{node}; if $0<\nu<2\sqrt{c(c-1)}$, the fixed point is unstable \emph{spiral}, see FIG. 6.
\end{itemize}

All the other remaining cases, i.e, both fixed points collide, are degenerate, since if $c=1$, $g(\lambda)=\lambda^2-\nu \lambda$.

Note that when there is no viscosity, the system has Hamiltonian, which is the Lyapunov function 
\begin{equation} \label{h}
{\mathcal H}(\mathbf{u}) =\frac{1}{2}\psi^2+\frac{1-c}{2c}\phi^2+\frac{1}{6c}\phi^3=\frac{1}{c}{\mathcal L}(\mathbf{u}),
\end{equation} 
see \eqref{e8}. Therefore, along any phase path ${\mathcal H}(\mathbf{u}) =constant$ is a conserved quantity, see FIG. 8. In this case, $p_0=p_1=0$, and hence
\begin{itemize}
\item[(i)] $(\phi^-,0)$ is a \emph{saddle} and $(\phi^+,0)$ is a \emph{center} when $q_1>0$, see FIG. 7 left,
\item[(ii)] $(\phi^-,0)$ is a \emph{center} and $(\phi^+,0)$ is a \emph{saddle} when $q_1<0$, see FIG. 7 middle, and
\item[(iii)] $c=1$ is degenerate since both critical points collide at origin $\phi^-=\phi^+$, see FIG. 7 right.  
\end{itemize}
Therefore, in this case the unstable spirals (which correspond to the case with small viscosity $\nu=0.1$) become centers when there is no viscosity $\nu=0$.
 
This is an example of a transcritical bifurcation where, at the intersection of the two bifurcation curves $\phi=0$ and $\phi=2(c-1)$, the equilibrium changes from one curve to the other at the bifurcation point. As $c$ increases through one,  the saddle point collides with the unstable node, and then remains there whilst the unstable node or spiral moves away from $(\phi^-,0)$.
\medskip


\section{Summary}

In this paper a basic theory for the BBM equation \eqref{BBM}, and its extension \eqref{e0} to include the dissipation term was shown. When the viscous term is not present, \eqref{BBM} has traveling wave solutions that depend critically on the traveling wave velocity. When the velocity $c>1$  \eqref{e0}  has solitary waves solutions, while if $c<1$, the solutions become periodic. At the interface between the two cases, when $c=1$, the solutions are rational functions ofcanoidal waves. Based on Ince's transformation, a theory including the dissipative term was presented, in which we have found more general solutions in terms of Weierstrass elliptic $\wp$ functions.  Using dynamical systems theory we have shown that the solutions of \eqref{e0} experience a transcritical bifurcation when $c=1$, where at the intersection of the two bifurcation curves, stable equilibrium changes from one curve to the other at the bifurcation point. As the velocity changes, the saddle point collides with the node at the origin, and then remains there, while the stable node moves away from the origin.  Also, the bifurcation curves and the fixed points of the dynamical system are functions of the traveling wave velocity $c$ and dissipation constant $\nu$.

\begin{figure*}[ht!]
\begin{center}
\begin{tabular}{lll}
\resizebox*{0.33\textwidth}{!}{\rotatebox{0}
{\includegraphics{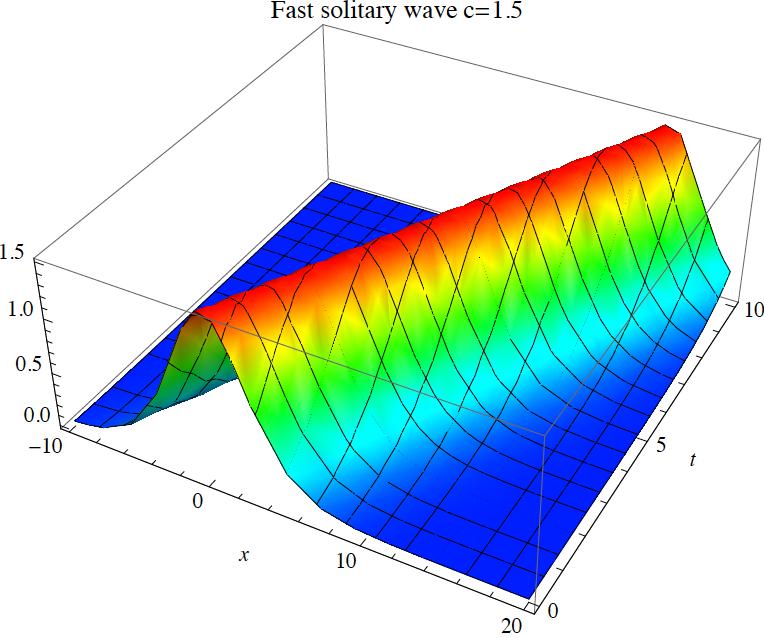}}}
&
\resizebox*{0.33\textwidth}{!}{\rotatebox{0}
{\includegraphics{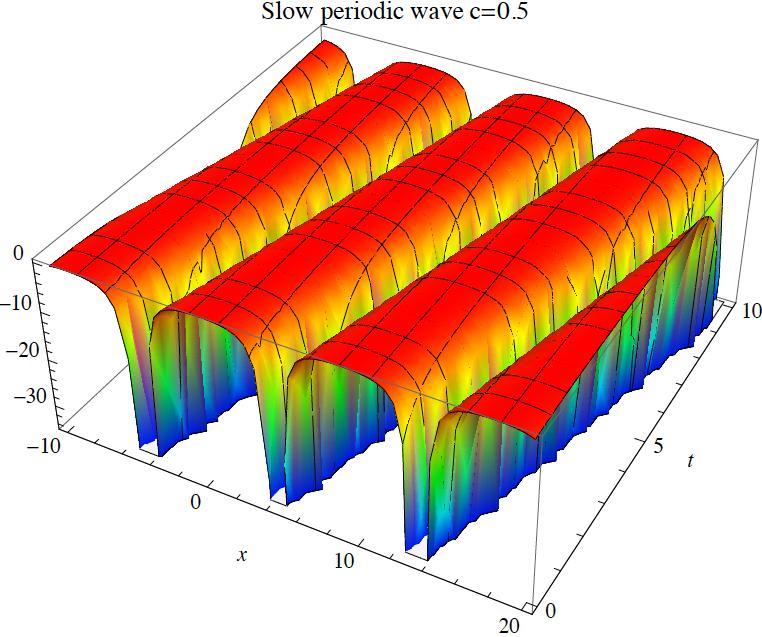}}}
&
\resizebox*{0.33\textwidth}{!}{\rotatebox{0}
{\includegraphics{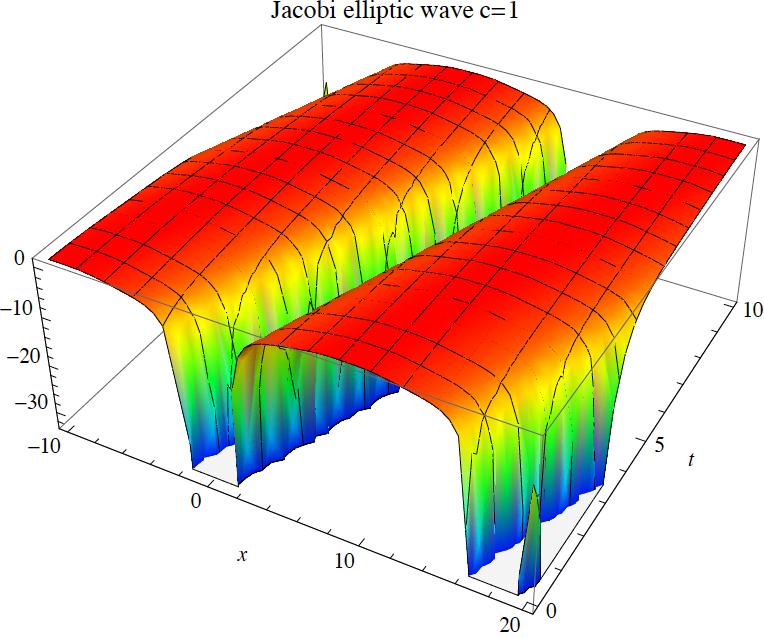}}}
\end{tabular}
\caption{Traveling waves $\nu=0$, left $c=1.5$ and middle $c=0.5$ eq. \eqref{62}; right  $c=1$ eq. \eqref{el8} }
\end{center}
\end{figure*}

\begin{figure*}[ht!]
\begin{center}
\begin{tabular}{lll}
\resizebox*{0.33\textwidth}{!}{\rotatebox{0}
{\includegraphics{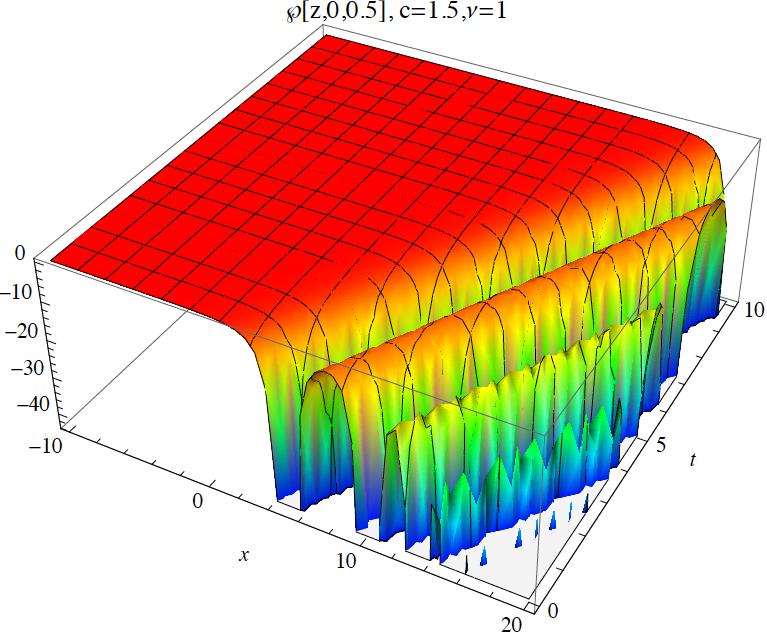}}}
&
\resizebox*{0.33\textwidth}{!}{\rotatebox{0}
{\includegraphics{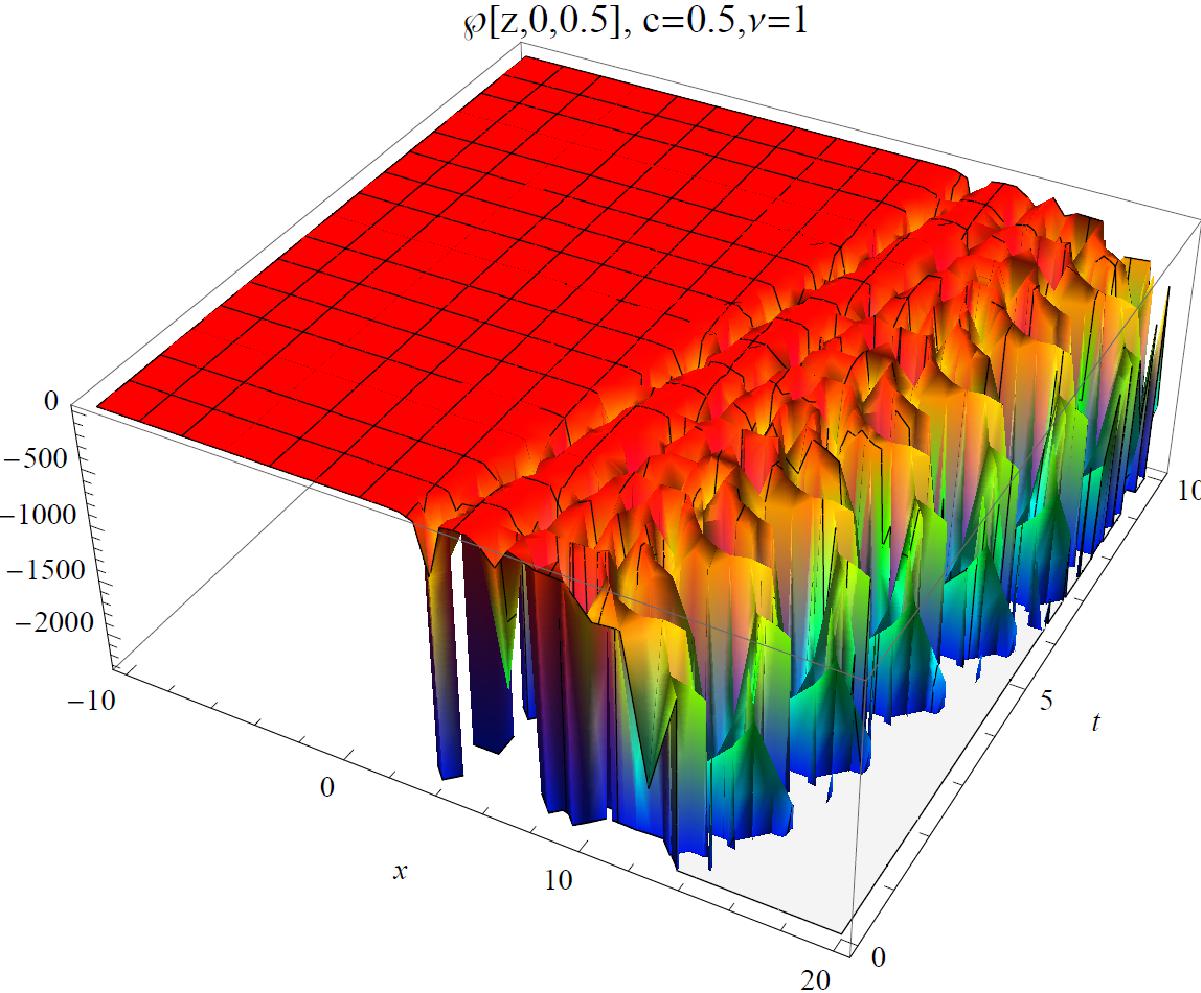}}}
&
\resizebox*{0.33\textwidth}{!}{\rotatebox{0}
{\includegraphics{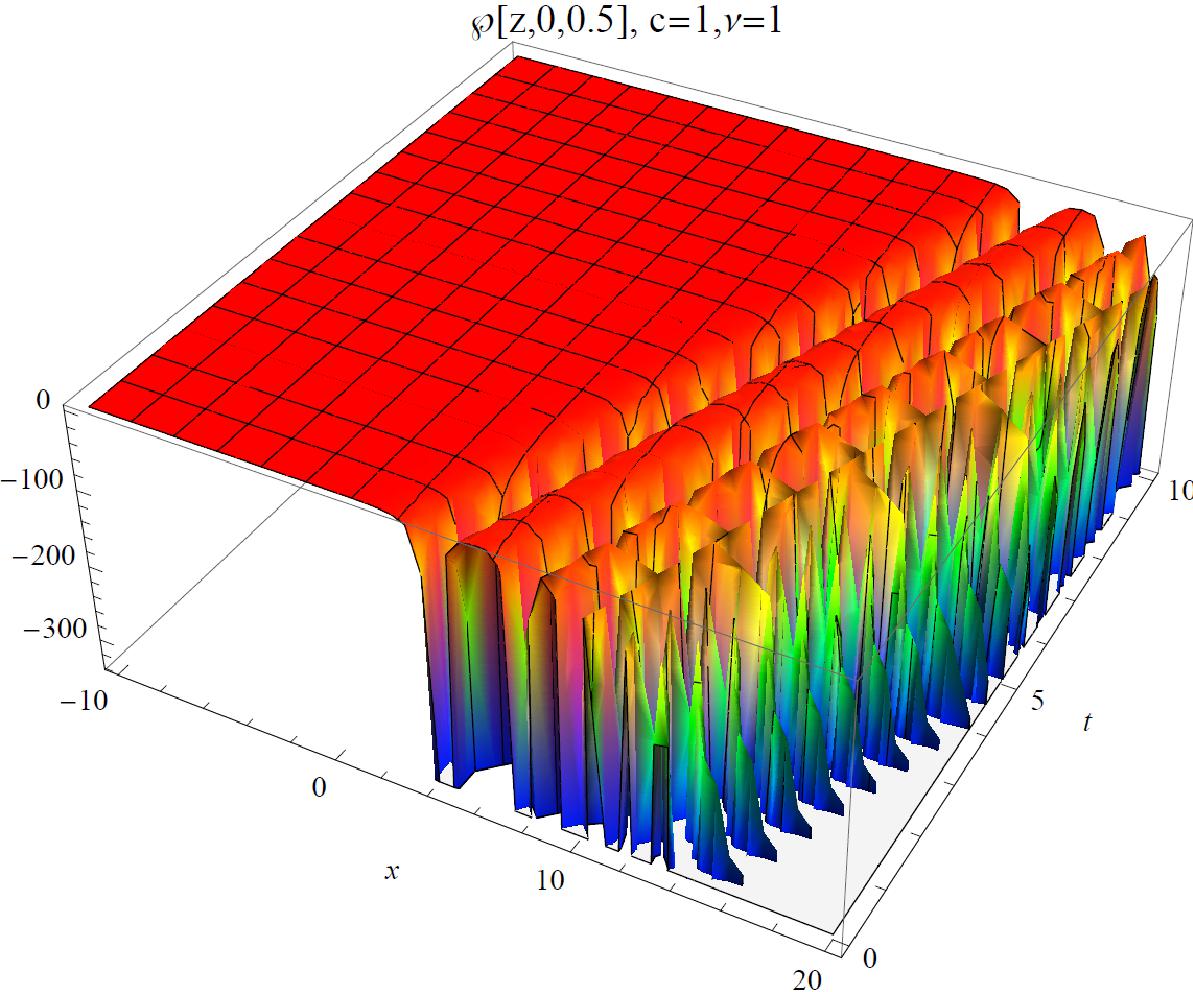}}}
\end{tabular}
\caption{Weierstrass solutions $\nu=1$ eq. \eqref{sol}, left $c=1.5$; middle $c=0.5$; right $c=1$}
\end{center}
\end{figure*}

\begin{figure*}[ht!]
\begin{center}
\begin{tabular}{lll}
\resizebox*{0.33\textwidth}{!}{\rotatebox{0}
{\includegraphics{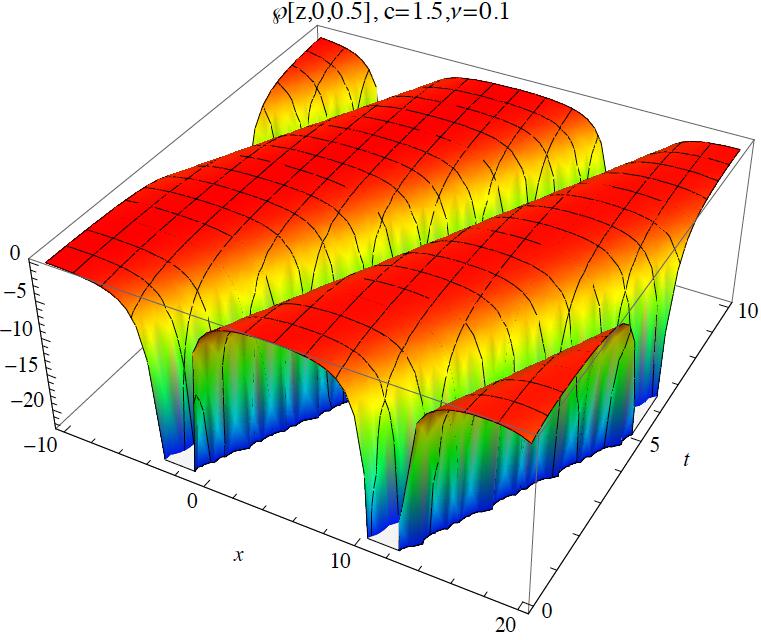}}}
&
\resizebox*{0.33\textwidth}{!}{\rotatebox{0}
{\includegraphics{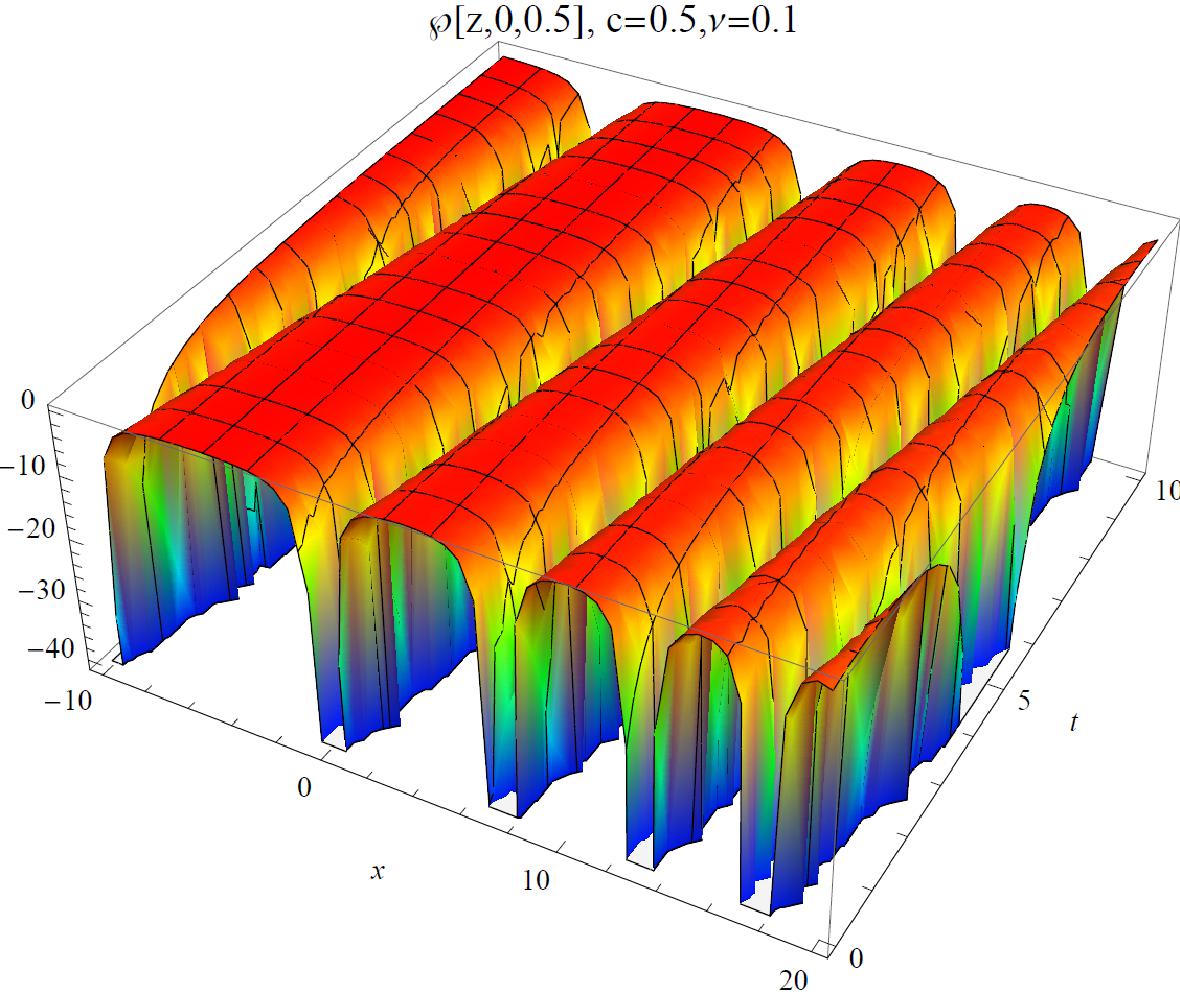}}}
&
\resizebox*{0.33\textwidth}{!}{\rotatebox{0}
{\includegraphics{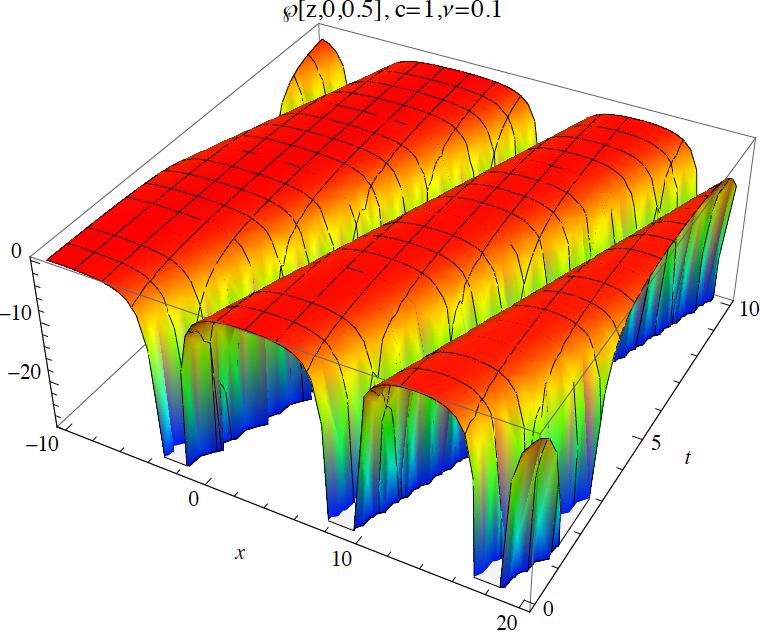}}}
\end{tabular}
\caption{Weierstrass solutions $\nu=0.1$ eq. \eqref{sol}, left $c=1.5$; middle $c=0.5$; right $c=1$}
\end{center}
\end{figure*}

\begin{figure*}[ht!]\label{figure4}
\begin{center}
\begin{tabular}{lll}
\resizebox*{0.33\textwidth}{!}{\rotatebox{0}
{\includegraphics{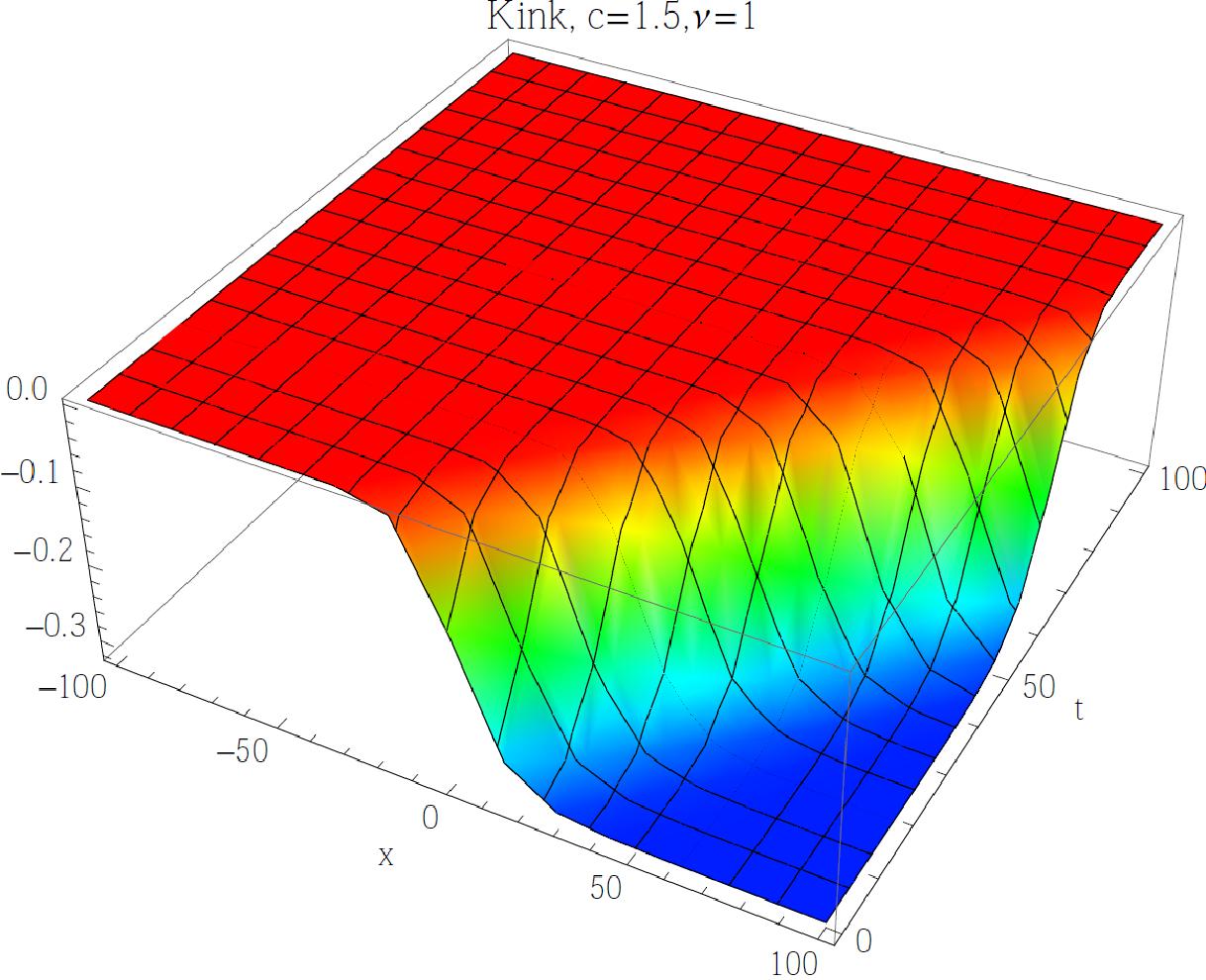}}}
&
\resizebox*{0.33\textwidth}{!}{\rotatebox{0}
{\includegraphics{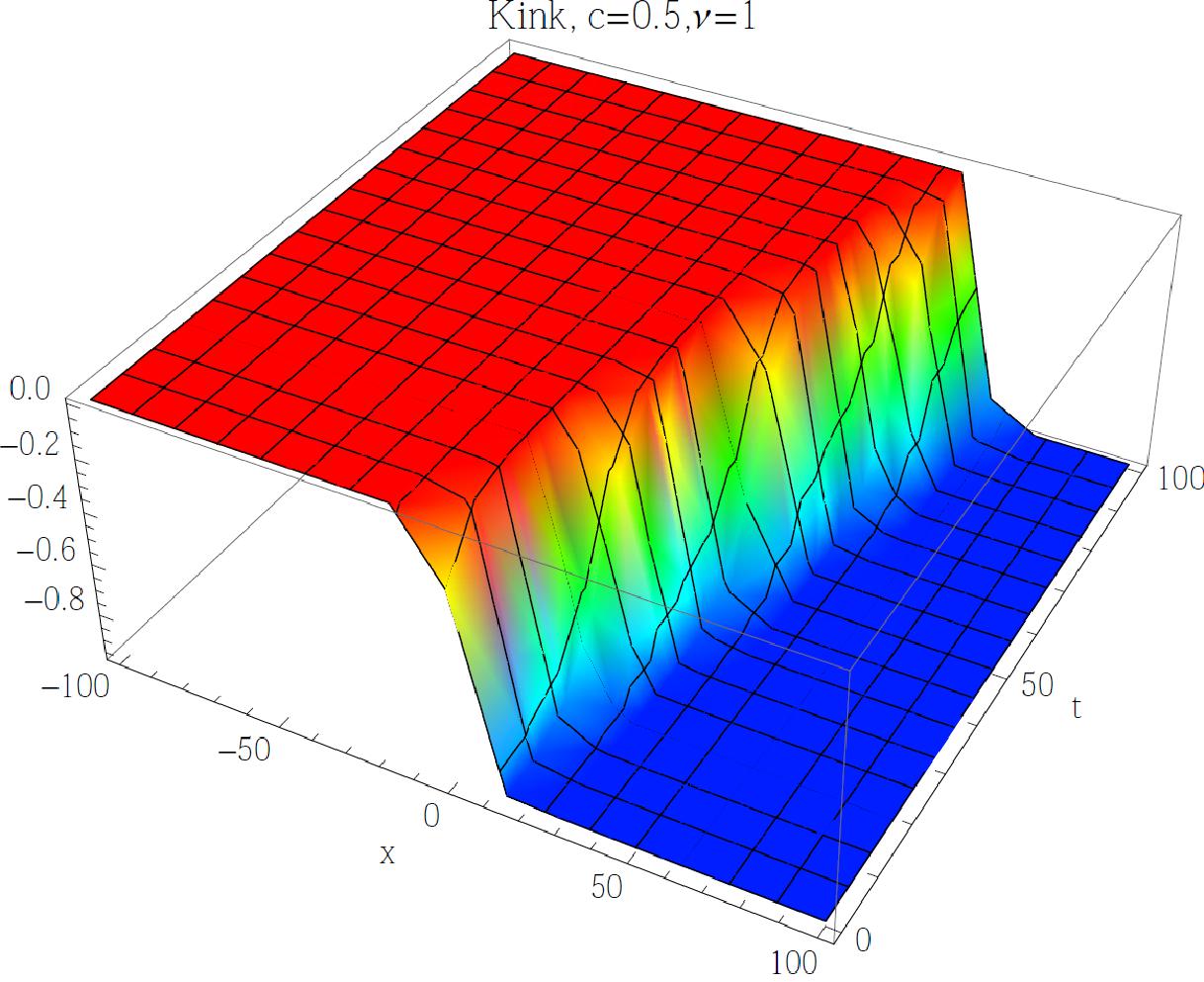}}}
&
\resizebox*{0.33\textwidth}{!}{\rotatebox{0}
{\includegraphics{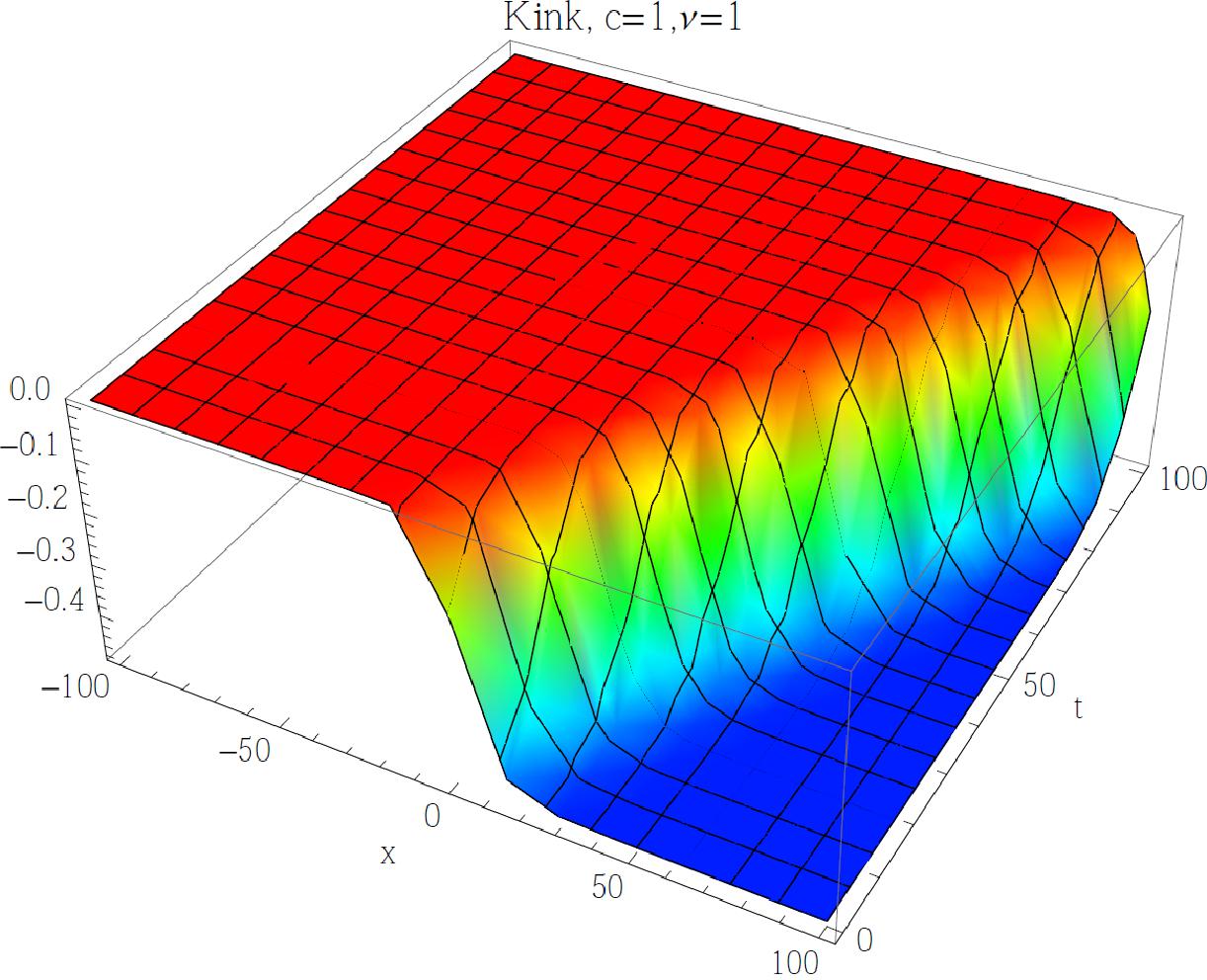}}}
\end{tabular}
\caption{Kink solutions $\nu=1$ eq. \eqref{112}, left $c=1.5$; middle $c=0.5$; right $c=1$ }
\end{center}
\end{figure*}

\begin{figure*}[ht!]\label{figure5}
\begin{center}
\begin{tabular}{ll}
\resizebox*{0.33\textwidth}{!}{\rotatebox{0}
{\includegraphics{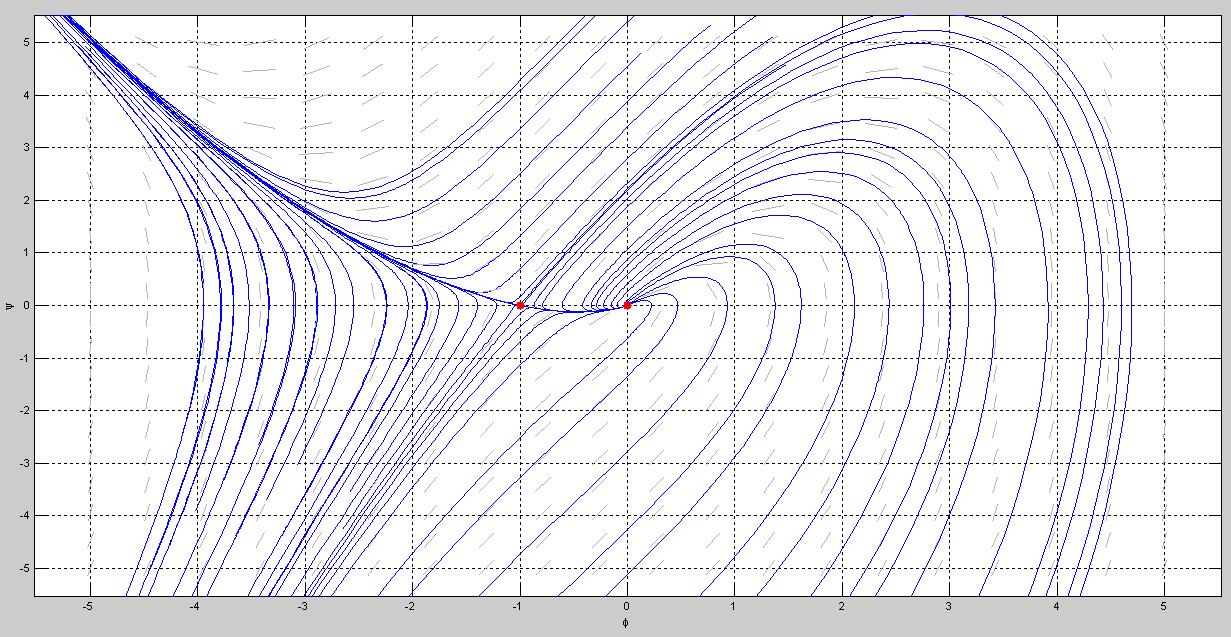}}}
&
\resizebox*{0.33\textwidth}{!}{\rotatebox{0}
{\includegraphics{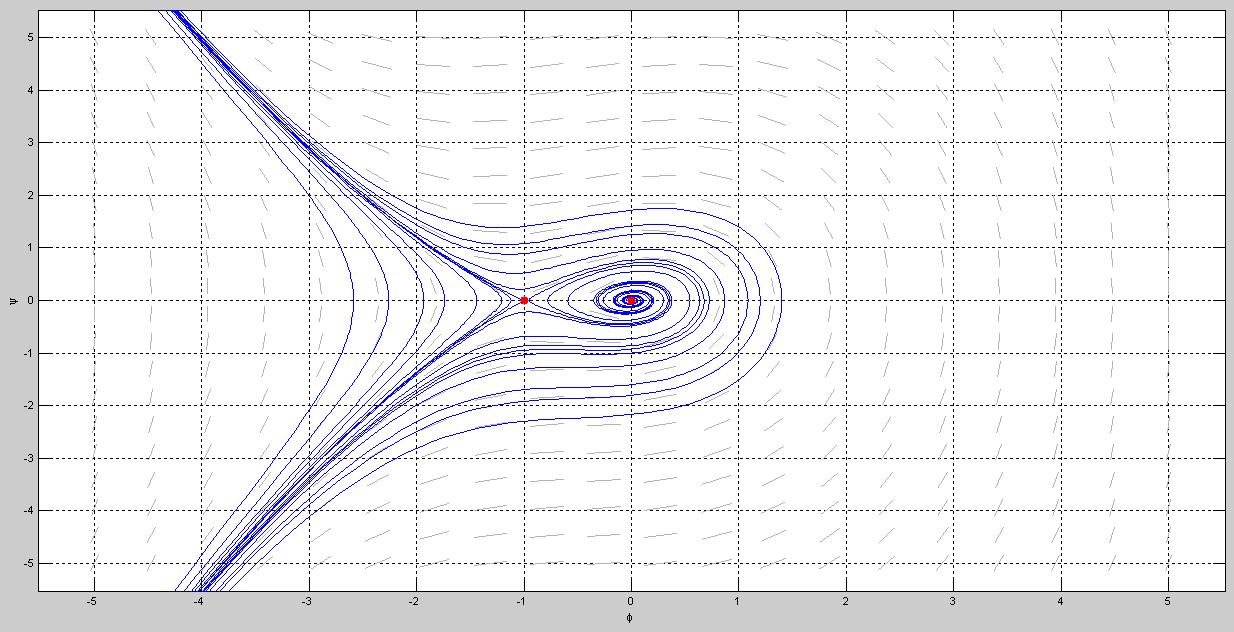}}}
\end{tabular}
\caption{Phase-plane for $c=0.5$, left $\nu=1, (0.0)$ node, $(-1,0)$ saddle; right $\nu=0.1, (0.0)$ spiral, $(-1,0)$ saddle}
\end{center}
\end{figure*}

\begin{figure*}[ht!]\label{figure6}
\begin{center}
\begin{tabular}{ll}
\resizebox*{0.33\textwidth}{!}{\rotatebox{0}
{\includegraphics{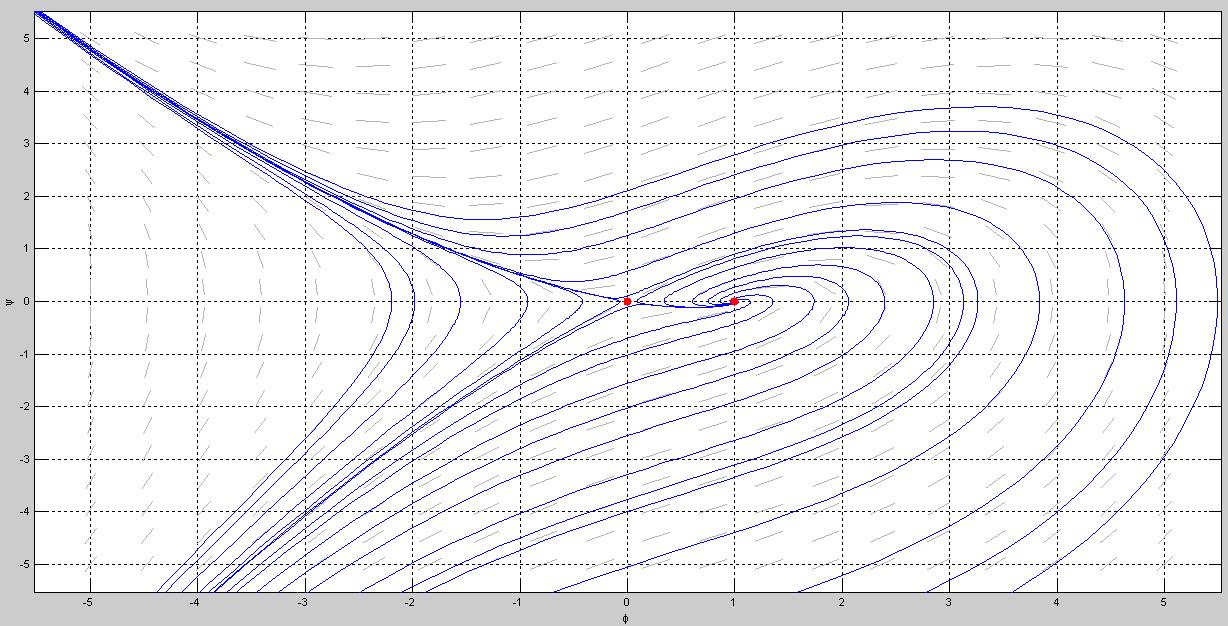}}}
&
\resizebox*{0.33\textwidth}{!}{\rotatebox{0}
{\includegraphics{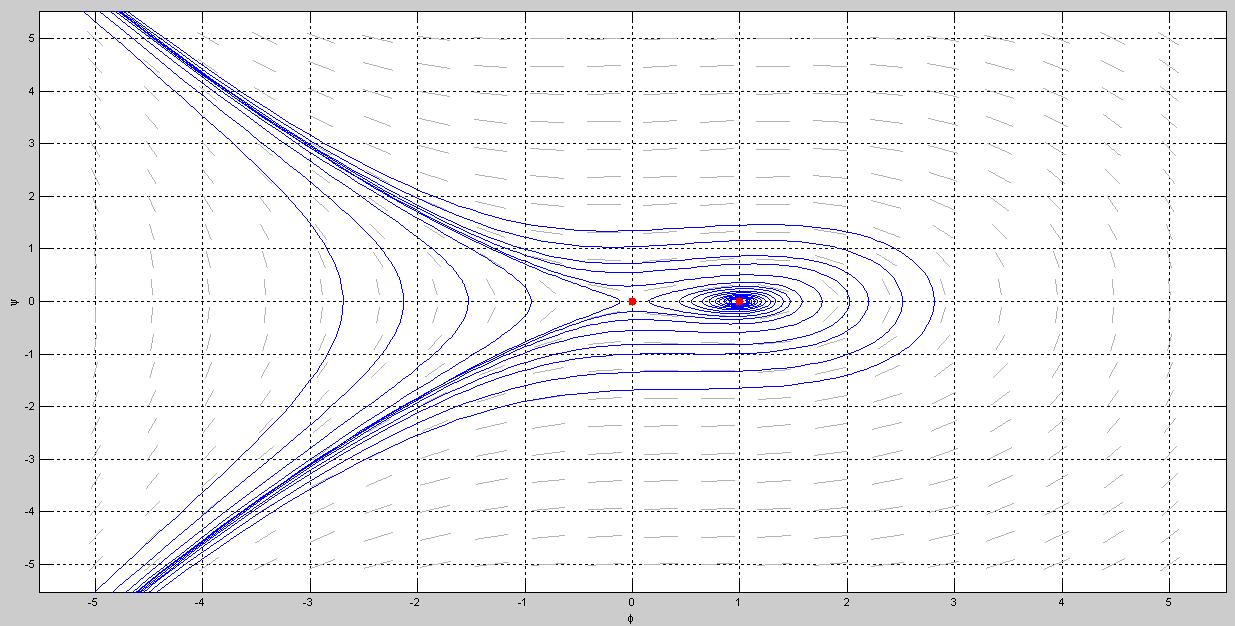}}}
\end{tabular}
\caption{Phase-plane for $c=1.5$, left $\nu=1, (0.0)$ saddle, $(1,0)$ node; right $\nu=0.1, (0.0)$ saddle, $(1,0)$ spiral }
\end{center}
\end{figure*}

\begin{figure*}[ht!]\label{figure7}
\begin{center}
\begin{tabular}{lll}
\resizebox*{0.33\textwidth}{!}{\rotatebox{0}
{\includegraphics{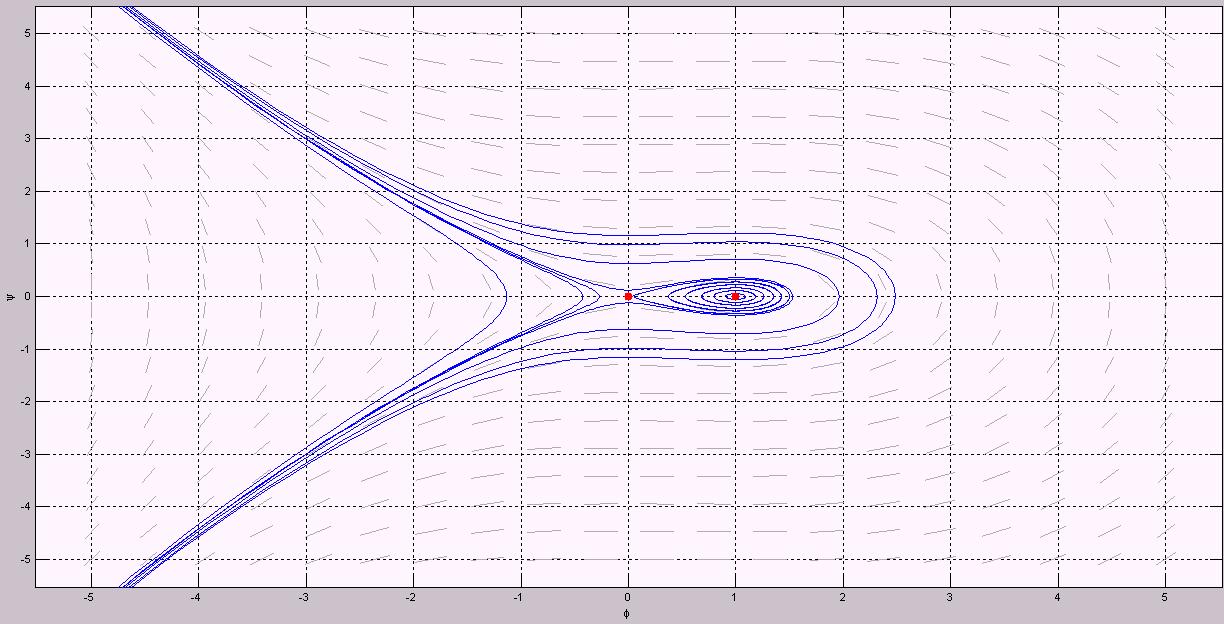}}}
&
\resizebox*{0.33\textwidth}{!}{\rotatebox{0}
{\includegraphics{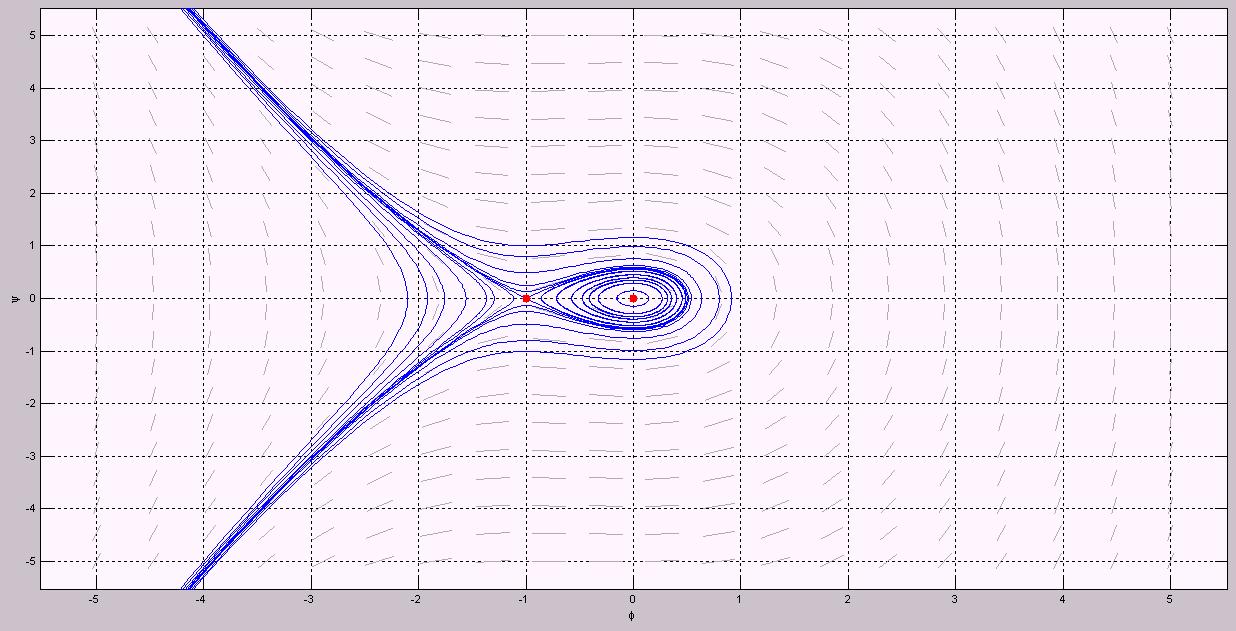}}}
&
\resizebox*{0.33\textwidth}{!}{\rotatebox{0}
{\includegraphics{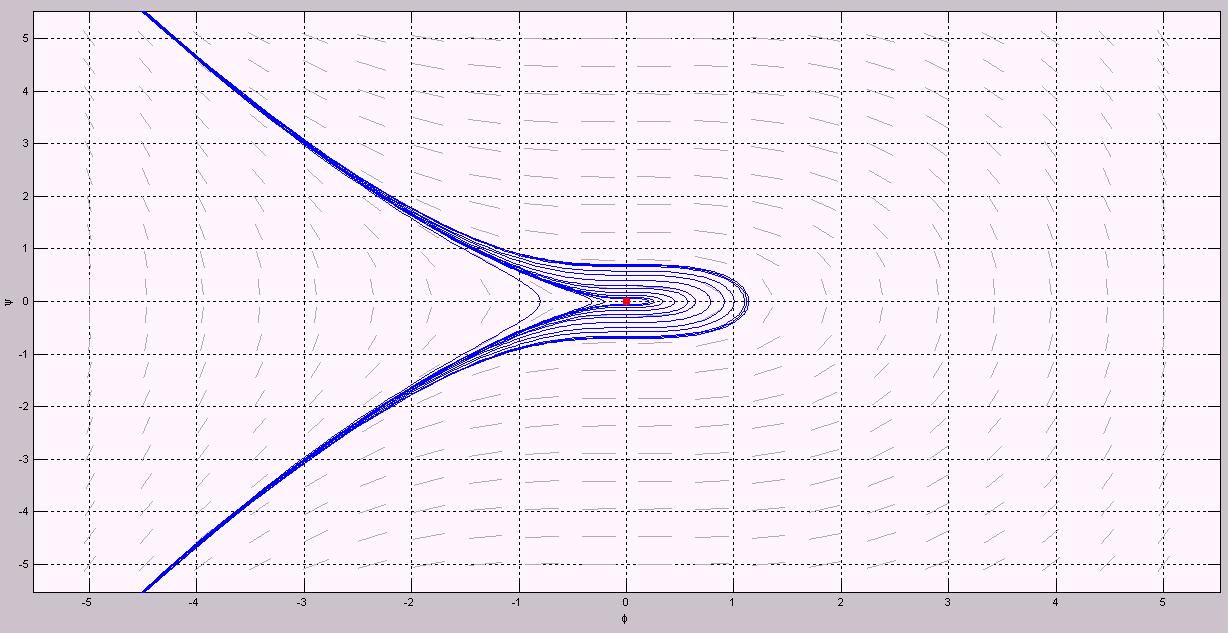}}}
\end{tabular}
\caption{Phase-plane for $\nu=0$, left $c=1.5, (0.0)$ saddle, $(1,0)$ center; middle $c=0.5$, $(0.0)$ center, $(-1,0)$ saddle; right $c=1$ $(0.0)$ degenerate}
\end{center}
\end{figure*} 

\begin{figure*}[ht!]\label{figure8}
\begin{center}
\begin{tabular}{lll}
\resizebox*{0.25\textwidth}{!}{\rotatebox{0}
{\includegraphics{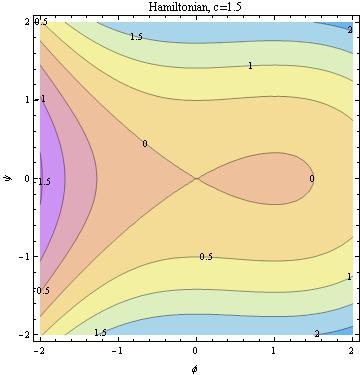}}}
&
\resizebox*{0.25\textwidth}{!}{\rotatebox{0}
{\includegraphics{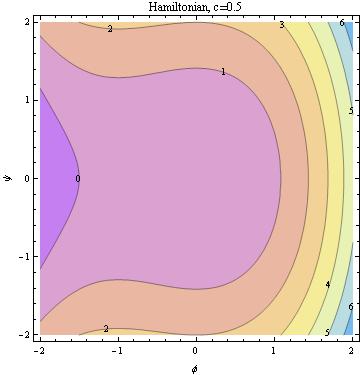}}}
&
\resizebox*{0.25\textwidth}{!}{\rotatebox{0}
{\includegraphics{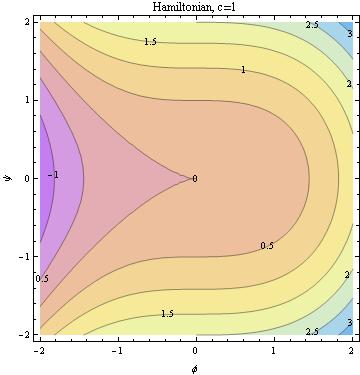}}}
\end{tabular}
\caption{Contour plot of Hamiltonian $\mathcal H=const$ see \eqref{h}, left $c=1.5$; middle $c=0.5$; right $c=1$}
\end{center}
\end{figure*}


\end{document}